\documentclass[apj]{emulateapj}

\slugcomment{}

\shorttitle{}
\shortauthors{Camporeale \& Burgess}

\begin{document}


\title{Electron temperature anisotropy in an expanding plasma: Particle-in-Cell simulations}


\author{E. Camporeale and D. Burgess}
\affil{Queen Mary University of London, Astronomy Unit, E1 4NS, London, UK}


\begin{abstract}
We perform fully-kinetic particle-in-cell simulations of an hot plasma that
expands radially in a cylindrical geometry. The aim of the paper is to study the consequent development of the electron temperature anisotropy in an expanding plasma flow as found in a collisionless stellar wind.
Kinetic plasma theory and simulations have shown that the electron temperature anisotropy is controlled by fluctuations driven by electromagnetic kinetic instabilities. In this study the temperature anisotropy is driven self-consistently by the expansion. While the expansion favors an increase of parallel anisotropy ($T_\parallel>T_\perp$), the onset of the firehose instability will tend to decrease it. We show the results for a supersonic, subsonic, and static expansion flows, and suggest possible applications of the results for the solar wind and other stellar winds.
\end{abstract}


\keywords{}



\section{Introduction}

The steady-state solution of the most simple dynamical equations that describe the evolution of a stellar
wind dates back to the hydrodynamic model for the solar wind proposed by Parker in 1958 \citep{parker58}.
Increasingly sophisticated models have been developed to take into account the different observed characteristics of the
solar wind, and modeling has been based on analytic, semi-analytic and simulational methods \citep{hollweg08}. The magnetohydrodynamics (MHD) approach is widely used for global modeling, but since it treats the plasma as a fluid it does not include any effects due to non-Maxwellian particle velocity distributions. More recently, kinetic simulations have tried to
take in account and explain features related
to the observed non-thermal distribution of particles in velocity space (see e.g. \citet{landi2003, zouganelis05} ). The aim of such exospheric models is to provide a realistic description of the wind dynamics that includes the transition from a collision-dominated to a collisionless region. In doing so, however these models do not include the effect of plasma
instabilities and therefore cannot be regarded as completely self-consistent.

For the solar wind, the importance of small scale fluctuations, associated with kinetic plasma instabilities generated by non-Maxwellian particle distributions, is now widely recognized. This has come about through a convergence of observations, theory and simulations.
It is argued that many macroscopic quantities that characterize the solar wind, such as the particle temperature anisotropy or the electron heat flux, are always observed with values that are bounded by the possible
onset of kinetic instabilities. The general argument is the following: the expansion of the flow leads to a distortion of the distribution function, which represents, for example, an increase in the temperature anisotropy. The distortion of the distribution function can be enough to trigger linear instability, i.e., there is some free-energy available which can create plasma fluctuations. Based on simulations of the initial value problem for such instabilities, it is evident that the fluctuations act to reduce the distortion of the distribution function. In other words, the primary effect of the fluctuations produced in the instability is to restore a stable (or marginally stable) distribution function. In the case of a temperature anisotropy driven by expansion, we might expect the onset of instabilities associated with temperature anisotropy to act as an upper limit for the anisotropy. This argument can be broadly applied to many situations, such as formation of heat flux, or indeed compression flows.

In the solar wind context it is expected that the
expansion of the wind away from the Sun produces a parallel temperature anisotropy $T_\parallel>T_\perp$ (with reference to the magnetic field direction) due to the conservation of adiabatic
invariants. If one adopts a fluid viewpoint (for instance the `double-adiabatic' theory \citep{CGL} that is often used in this context), the anisotropy is not bounded and it should increase indefinitely
for increasing distance from the Sun. However it has been observed that the temperature anisotropy is limited in value both for ions \citep{gary01,kasper02,hellinger06,matteini07} and
electrons \citep{gary05,stverak08}, and it has been suggested that
this is due to the onset of kinetic instabilities, such as the firehose for $T_\parallel>T_\perp$ and the mirror or ion-cyclotron instabilities for $T_\parallel<T_\perp$. Recently, several computer simulations have been able
to show that the kinetic instabilities are effectively able to
constrain the growth of temperature anisotropy.
Those include hybrid simulations for protons \citep{hellinger03,matteini06} and particle-in-cell (PIC) simulations for electrons \citep{gary96,gary00,gary03,camporeale08}.

Most simulations study the initial value problem for an instability, starting with some initially unstable anisotropy value and following the evolution of the system to marginal stability. These simulations are usually interpreted in the framework of the linear theory of the Vlasov-Maxwell equations. The linear theory is developed for a non-expanding plasma embedded in a uniform magnetic field, in a periodic Cartesian geometry. It is this theory which is used to determine the marginal stability boundary used to confirm the role of linear instabilities as constraining the observed particle anisotropy. In other words, the paucity of observed periods of large anisotropy is interpreted as a consequence of fluctuations driven by anisotropy instabilities which stop the plasma reaching a high anisotropy state, or rapidly relaxing it if it does. The bounds of the anisotropy are calculated using a non-expanding Cartesian geometry.

All the simulations performed so far do not include the radial expansion of the wind in a completely consistent manner. Because of the computational difficulties of simulating an expanding plasma in a fixed frame of reference, simulations using an `expanding box model' have been
developed. In this type of simulation the idea is to use a Cartesian computational domain, but to `stretch out' the domain as the plasma expands, thus distorting the simulation box in at least one dimension. As a consequence, the equations of motion have to be modified, taking into account the inertial forces due to the expansion of the box, and the coordinates must be continuously rescaled. For plasma simulations this method was initiated in \citet{grappin93} for MHD, and it has been successively applied to hybrid simulations (with fluid electrons and particle ions) in \citet{hellinger03}. Although the simulation results have been relatively successful, and shown to be consistent with satellite observations \citep{matteini07}, the expanding box model is still unable to treat the expansion consistently. It can be argued that the rescaling of the box effectively produces an unphysical mode coupling due to the allowed wave vectors changing continuously over time. Energy that resides in certain modes at one time is artificially channelled towards other modes, as time evolves. The counter argument is that this mode evolution is actually expected as precisely  a result of the expansion.\\
In this paper we present the first fully-kinetic PIC simulations, with realistic proton-electron mass ratio, of a plasma that expands radially in a decreasing magnetic field, in a fixed frame of reference. The expansion is radial in a cylindrical geometry. The scope of this work is to study and quantify in a more consistent way the competition between the growth of parallel anisotropy due to the expansion, and the possible onset of kinetic instabilities. The simulations presented in this paper are 2D-3V, i.e. two dimensional in space and three dimensional in velocity space. The magnetic field is also 2D, with axial component zero. Clearly, this is just a first step towards more challenging, and realistic 3D simulations. But, we will show that even with the use of cylindrical geometry it is already possible to capture the increase of anisotropy and the subsequent development of kinetic instabilities. With the scales currently feasible, our results are relevant to the evolution of the electron particle distribution. We will show and compare three different cases: static, subsonic, and supersonic flow.\\
The paper is organised as follows. The feasibility of running the simulations has been greatly enhanced by using an implicit scheme, and we discuss the main features of the algorithm in Section 2. Section 3 is devoted to describe the details of the box geometry and size, and the characteristic plasma timescale and lengths. We show and discuss the results of different runs in Section 4, and draw our conclusions in Section 5.

\section{The implicit code}
The code used in this work is a fully-kinetic, implicit, parallel, electromagnetic PIC code called \emph{PARSEK}. Its features are described in detail in \citet{markidis09}, but for completeness we give here a brief description of the algorithm. We refer the reader to, e.g., \citet{pritchett03} or \citet{hockney_book} for a more general tutorial on PIC techniques.\\
The algorithm is implicit in time both for the particle mover and for the field solver, and it adopts the so-called `implicit moment method', introduced by \citet{brackbill82}, and successively re-elaborated in \citet{vu92}, and \citet{ricci02}. The following equations (in CGS units) for the conservation of density and momentum are used to extrapolate the values of the charge density $\rho$, and the current density $\mathbf{J}$ at a new timestep:
\begin{equation}\label{0moment}
\frac{\rho^{i+1}-\rho^i}{\Delta t}+\nabla\cdot\mathbf{J}^{i+\frac{1}{2}}=0
\end{equation}
\begin{equation}\label{1moment}
\frac{\mathbf{J}^{i+1}-\mathbf{J}^i}{\Delta t}=\frac{q}{m}(\rho^i\mathbf{E}^{i+1}+\frac{\mathbf{J}^{i+\frac{1}{2}}\times\mathbf{B}^i}{c}-\nabla\cdot\mathbf{P}^i),
\end{equation}
where superscripts indicate timestep, and all other symbols are standard. The closure of equations (\ref{0moment}-\ref{1moment}) is provided by approximating the divergence of the pressure tensor at time $i+1$ with the value at time $i$, that is $\nabla\cdot\mathbf{P}^{i+1}\sim\nabla\cdot\mathbf{P}^{i}$. From Eqs. (\ref{0moment}-\ref{1moment}) one can formulate $\rho^{i+1}$ and $\mathbf{J}^{i+1}$ as functions of the electric field $\mathbf{E}^{i+1}$. By using those relations in Maxwell equations, and after some algebra one ends up with a linear equation for $\mathbf{E}^{i+1}$ as a function of only old quantities:
\begin{eqnarray}\label{GMRES}
(c\theta\Delta t)^2[-\nabla^2\mathbf{E}^{i+1}-\nabla\nabla\cdot(\mu^i\mathbf{E}^{i+1})]+(\mathbf{I}+\mu^i)\mathbf{E}^{i+1}=
\mathbf{E}^i+(c\Delta t)\Bigl(\nabla\times\mathbf{B}^i-\frac{4\pi}{c}\mathbf{\hat{J}^i}\Bigr)-(c\Delta t)^2\nabla 4\pi\hat{\rho}^i
\end{eqnarray}
where the following terms are defined:
\begin{eqnarray}
\mathbf{\hat{J}}^i&=&\sum_{s}\Pi^i\cdot[\mathbf{J}^i-\frac{q\Delta t}{2m}\nabla\cdot\mathbf{P}^i]\\
\Pi^i&=&[\mathbf{I}+\frac{q\Delta t}{2mc}\mathbf{I}\times\mathbf{B}^i+\frac{q^2\Delta t^2}{4m^2c^2} (\mathbf{I}\cdot\mathbf{B}^i)\mathbf{B}^i]/[1+\frac{q^2\Delta t^2 {B^i}^2}{4m^2c^2}]\\
\mu^i&=&\frac{\Delta t^2}{2} \omega_{ps}\Pi^i \;\;\;,\;\;\;\omega_{ps}=\frac{4\pi q\rho^i}{m}\\
\hat{\rho}&=&\rho-(\Delta t)\nabla\cdot\mathbf{\hat{J}}
\end{eqnarray}
and the subscript $s$ indicates the species.\\
Equation (\ref{GMRES}) is solved by a matrix-free Generalized Minimal Residual (GMRes) iterative linear solver \citep{saad86}. Once the electric field is known at time $i+1$, the magnetic field $\mathbf{B}$ is advanced using Faraday's law:
\begin{equation}
\mathbf{B}^{i+1}=\mathbf{B}^i-c\Delta t\nabla\times\mathbf{E}^{i+i}.
\end{equation}
The particle mover pushes the particles to a new position and velocity, according to the equations:
\begin{eqnarray}
\label{x}\mathbf{x}^{i+1}&=&\mathbf{x}^i+\mathbf{v}^{i+\frac{1}{2}}\Delta
t\\
\label{v}\mathbf{v}^{i+1}&=&\mathbf{v}^i+\frac{q\Delta
t}{m}\left(\mathbf{E}^{i+1}+\frac{\mathbf{v}^{i+\frac{1}{2}}\times\mathbf{B}^{i+1}}{c}\right).
\end{eqnarray}
Equations (\ref{x},\ref{v}) are also solved iteratively with a Predictor-Corrector technique. Finally, to ensure that the continuity equation is satisfied, the electric field must be corrected with:
\begin{equation}
\mathbf{E}_{new}=\mathbf{E}_{old}-\nabla\phi\;\;\;,\;\;\;\nabla^2\phi=\nabla\cdot\mathbf{E}_{old}-4\pi\rho.
\end{equation}
The main advantage of the implicit method over an explicit one is that it enables a choice of timestep $\Delta t$ and a grid size $\Delta x$ that do not satisfy the Courant stability condition $c\Delta t/\Delta x<1$. This is instead replaced by the less stringent accuracy condition $v_{e}\Delta t/\Delta x<1$, where $v_{e}$ is the electron thermal velocity. The timestep is still small enough to resolve the electron gyromotion. Of course this benefit is paid for in terms of the computational complexity of the algorithm, however for many situations the possibility of choosing a fairly large timestep results in a positive payoff for the total simulation runtime.

\section{Simulation setup}
\begin{figure}
\center
\includegraphics[width=80mm]{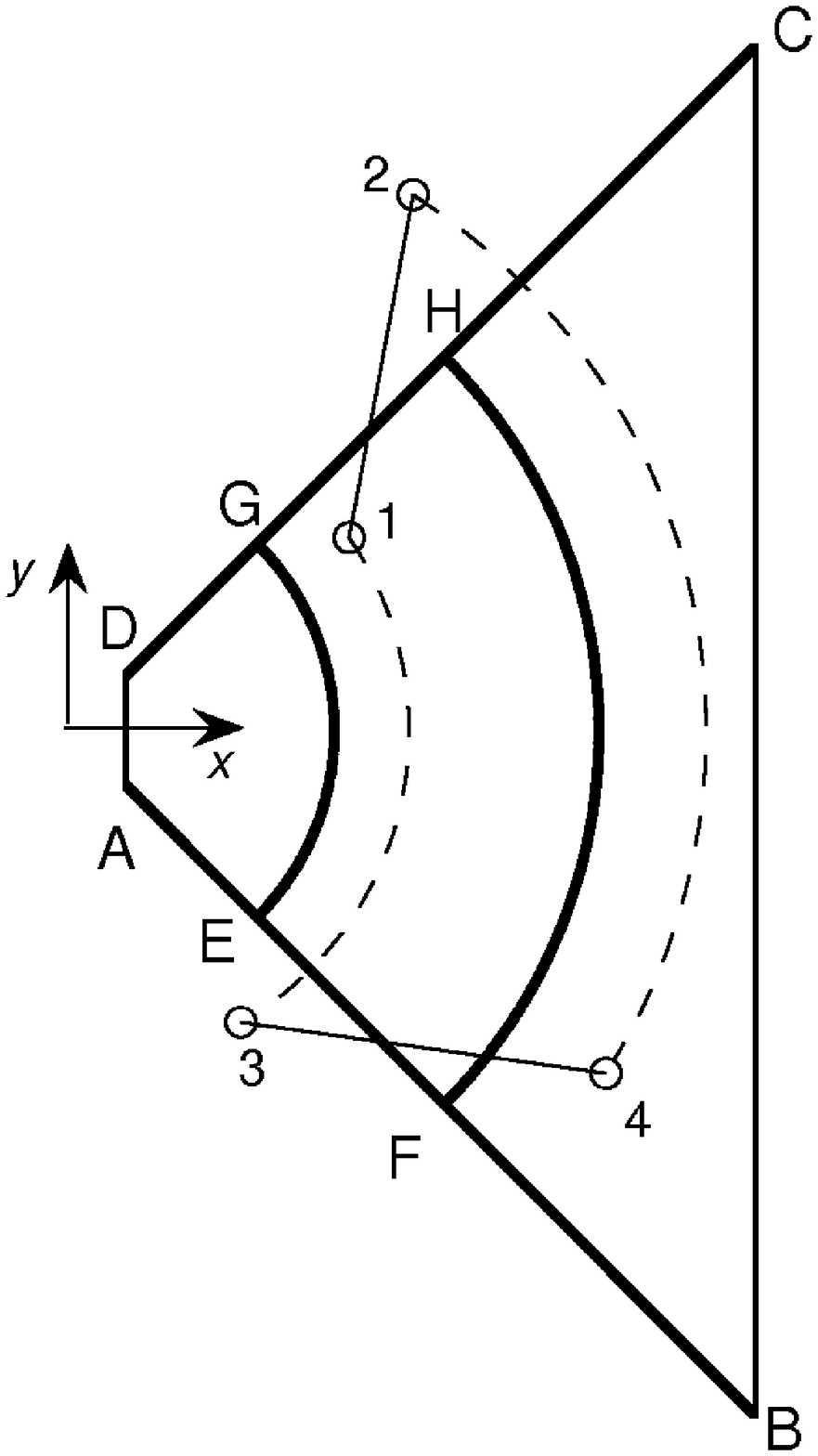}
\caption{Geometry of the computational box: periodic boundary conditions are applied on sides $AB$ and $CD$. $AEGD$ is the inner boundary region; $EFGH$ is the active region, and $FBCH$ is the outer boundary region. Particles are reinjected, with Maxwellian distribution, from sides $AD$ and $BC$.}\label{box}
\end{figure}

We simulate an ion-electron plasma, with physical mass-ratio, i.e., $m_i/m_e= 1836$. The plasma expands radially on a 2D disc, and the magnetic field is forced to have zero axial component so it is a 2D vector field. The geometry of the box is shown in Figure \ref{box}. The grid is Cartesian $(x,y)$ and covers the trapezoid $ABCD$. The oblique sides $AB$ and $DC$ form a $90^\circ$ angle, and we apply periodic boundary conditions on these sides. Therefore, a complete plane geometry is recovered by applying three successive 90 degrees rotations of the box. This four-fold symmetry is used to reduce the computational effort, and does not affect the short wave length fluctuations that develop.

To ensure a correct periodicity along the azimuthal direction, the particles that escape the boundary $DC$ are re-injected from the boundary $AB$, and vice versa. In doing so, their trajectory and velocities must be appropriately rotated by $90^\circ$. We show an example, in Figure \ref{box}, where a particle moving from point 1 to point 2 is re-injected at point 4, as if it was coming from point 3. The vector velocity $(v_x,v_y)$ on the plane must also be rotated. The same argument applies to the electric and magnetic fields on the boundary, where the $x$-component on the $AB$ side is imposed to match the $y$-component on the $DC$ side, while the $y$-component on $AB$ must be equal to the $x$-component on $AB$, with a change of sign (i.e. $E_y$ on $AB$ is equal to $-E_x$ on $DC$). The out-of-plane $z$ component is treated as periodic in the standard way.

We define three different regions in the box: an inner boundary region $AEGD$, an active region $EFHG$, and an outer boundary region $FBCH$. Particles are initially loaded in all the three regions, with a density that varies as $1/r$, with $r$ the distance from the origin $(0,0)$ (which is out of the box).
Except for the static case, they are initialised with an isotropic Maxwellian distribution, and with a radial mean velocity $V_m$.
The initial magnetic field on the plane is also radial, pointing outwards, and it decreases in magnitude as $1/r$. The initial electric field is null.
While particles are allowed to move anywhere in the box $ABCD$, the field solver advances the fields only in the active region.
This effectively produces boundary conditions on the arcs $EG$ and $FH$, where any perturbation of the initial fields is forced to smooth out.

The fact that we have a reservoir of particles in the inner and outer boundary regions that move consistently with a static electromagnetic field
avoids spurious boundary effects on the sides $EG$ and $FH$. We are therefore mainly interested in what happens inside the active region, which has consistent boundary conditions for both particles and fields on all its sides. The true boundaries of the computational box are of course the sides $AD$ and $BC$, which are sufficiently far from the active region. Here, as we said, the electromagnetic field is static; particles are allowed to escape, and they are re-injected at every timestep on both sides. The re-injection routine is computationally expensive, but it mimics the existence of a population of Maxwellian particles outside the box, with drift velocity equal to $V_m$ parallel to the magnetic field, and with density that again goes as $1/r$.

The plasma beta for the species $s$ is defined as usual as
$\beta_s=\frac{8\pi nT_s}{\mathbf{B}^2}$, where $n$ is the
density, $T$ the temperature, and $\mathbf{B}$ the magnetic field.
It increases linearly with the distance from the
origin, and the initial values of $n,T,\mathbf{B}$ are chosen so
that $\beta=7.7$ on the arc $EG$, and $\beta=15.4$ on $FH$, for
both electrons and ions (that is $T_e=T_i$). These relatively high
values of beta allow the expanding plasma to reach quickly a
parallel anisotropy large enough to trigger the electron firehose
instability. According to the linear kinetic theory, the only two parameters that control the growth rate of an anisotropy instability are the parallel beta $\beta_{\parallel}$ and the
temperature ratio $T_\perp/T_\parallel$. For the electrons,
\citet{camporeale08} have found that the relationship
\begin{equation}\label{thre}
\frac{T_\perp}{T_\parallel}=1-\frac{1.29}{\beta_{\parallel
e}^{0.98}}
\end{equation}
 is valid at the threshold of the instability.

Velocities are normalized to the light speed $c$, and the initial thermal velocities for electrons and ions are respectively $v_e=8\cdot10^{-3}$, and $v_i=1.9\cdot10^{-4}$.
The subsonic and supersonic case that we discuss in the next section are respectively for $V_m/v_e=0.625$, and $V_m/v_e=2.5$, while for the static case $V_m=0$, but the initial electron anisotropy is $T_\perp/T_\parallel=0.7$. The box has a maximum of 540 cells in the $x$ direction and 1170 in the $y$ direction. The active region has a radial extension of $L\omega_i/c=1.25$, where $\omega_i$ is the ion plasma frequency. The total box has a maximum radial extension of $4.22c/\omega_i$. The box is therefore sufficiently large to capture waves with wavenumber $kc/\omega_i$ greater than about 5. The timestep is $\Delta t= 0.05 \omega_i^{-1}$, and the cell size is $\Delta x=\Delta
y= 0.0055 c\omega_i^{-1}$, making the Courant parameter $c\Delta t/\Delta x$ equal to 9; the advantage of the implicit scheme is evident.

\section{Results}
In the CGL `double adiabatic' description of a plasma, the
quantities $\frac{T_\perp}{\mathbf{B}}$ and $\frac{T_\parallel\mathbf{B}^2}{n^2}$ are constant. It follows that the
temperature anisotropy varies as $T_\parallel/T_\perp \sim n^2/\mathbf{B}^3$. In
our configuration, where density and magnetic field decrease as
$1/r$, $T_\parallel/T_\perp$ will grow linearly with the distance.
Since we start with an isotropic particle distribution, the anisotropy is
expected to increase until the local plasma parameters are such
that Eq. (\ref{thre}) holds. At this point the electron firehose instability is triggered.

The linear dispersion relation of the electron firehose instability yields
two solutions \citep{li00}. One branch is a propagating slowly
growing mode, with angle of propagation ranging from $0^\circ$ to about $70^\circ$, and the other is a non-propagating fast growing mode, with wavevector forming an angle between about $30^\circ$ and $90^\circ$ with
the magnetic field. The latter is generally
dominant, but it has been shown that depending on the angle of
propagation and the level of anisotropy, there are wavevectors for
which the growth rate of the two modes is comparable \citep{camporeale08}.

\subsection{Supersonic case: $V_m/v_e=2.5$}
\begin{figure}
\center
\includegraphics[width=90mm]{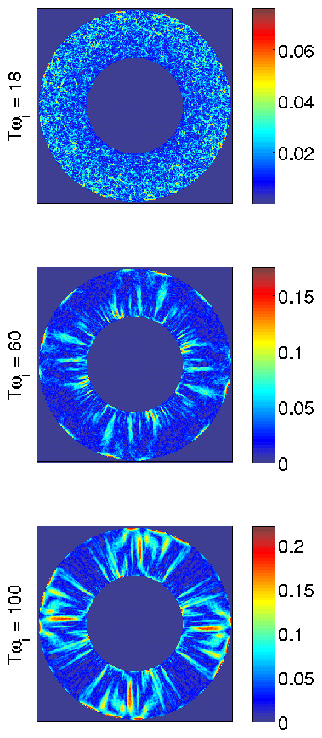}
\caption{Supersonic case: $\delta \mathbf{B}/\mathbf{B_0}$ at times $T\omega_i=18$ (top), $T\omega_i=60$ (middle), $T\omega_i=100$ (bottom). The images are obtained by successive rotations of the box.}\label{delta_B_super}
\end{figure}

In order to evaluate the role of instabilities in reducing the
electron temperature anisotropy in the expanding plasma, we
compare in this section two simulations. One has the setup described in the previous section, i.e the box is divided
in three regions, and the electromagnetic (EM) field is solved only in the active region. The second simulation is identical in all the parameters,
except for the EM field that is kept static through all the box. In this way, the differences between the two runs are clearly due to feedback effects caused by self-consistently generated electromagnetic fluctuations. We will call the two simulations `self-consistent' and `test-particle' runs. The focus of our interest will be the development of electron parallel anisotropy. Firstly, however, we show in Figure \ref{delta_B_super} the development at three successive times of the amplitude of the magnetic field fluctuations $\delta \mathbf{B}/\mathbf{B_0}$
 (where $\delta \mathbf{B}=|\mathbf{B}-\mathbf{B_0}|$, and
 $\mathbf{B_0}$ is the initial magnetic field), for the self-consistent run. These images have been obtained by successive rotations of the box. At time
 $T\omega_i=18$ (top panel), there is not yet any large scale
structure evident, and at that time there are many modes at different orientation but comparable amplitude present, creating an
almost random patchwork of magnetic field fluctuations. As time
evolves (middle and bottom panels), a structure of fluctuations aligned with the background  magnetic field emerges. This is consistent with the
 development of quasi-perpendicular waves, that at those times have superseded the more parallel modes. By performing a Fourier transform in time, we have indeed confirmed that the waves observed in Figure \ref{delta_B_super} are non-propagating (or alternatively moving very slowly, compared to the total time of the simulation) in the azimuthal direction. Movies of the evolution show that the structures do not propagate azimuthally, but do have features indicating that there is underlying convection outwards in the radial direction.

This is a first evidence that a plasma that increases its electron temperature anisotropy in a self-consistent expansion actually triggers a non-propagating firehose instability. This is not a trivial result, because although predicted by the
linear kinetic theory in Cartesian geometry, this has never been confirmed before by computer simulations, for an expanding plasma. As we discussed in the Introduction, what was already known is that an instability would develop if the plasma was starting with a sufficiently anisotropic temperature \citep{gary03,camporeale08}, while here we started with an isotropic plasma and let the anisotropy grow self-consistently, due to the
expansion. The physical behaviour here is much more complex than for a non-flowing plasma in a constant magnetic field. The magnetic fluctuations created by the instability are now convected outwards with the flow. This effect helps to lower further the temperature anisotropy at larger distances, by increasing the particle scattering.

\begin{figure}
\center
\includegraphics[width=80mm]{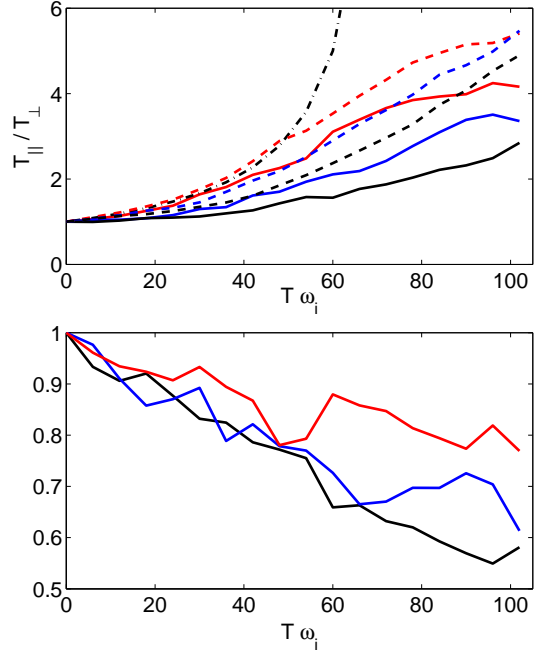}
\caption{Supersonic case. Top panel: anisotropy $T_\parallel/T_\perp$ at radial distances: $r\omega_i/c=1.50$ (red), $r\omega_i/c=1.86$ (blue), $r\omega_i/c=2.22$ (black). Solid lines are for the self-consistent run, and dashed lines are for  the test-particle run. The dot-dashed line is the CGL prediction for $r\omega_i/c=1.50$. Bottom panel: ratio of anisotropy for self-consistent over test-particle runs.}\label{anisotropy_super}
\end{figure}

The top panel of Figure \ref{anisotropy_super} shows the development of the anisotropy at three different radial distances, where the temperatures are
averaged over the azimuthal direction. The three solid
lines are for $r=1.50$ (red), $r=1.86$ (blue),
and $r=2.22$ (black), for the self-consistent run ($r$ is normalized to the ion inertial length $c\omega_i^{-1}$). The corresponding dashed lines show the development of the anisotropy at the same radial distances, for the test-particle simulation. We have also included, for
 $r=1.50$ only, the value of anisotropy predicted by the CGL theory (dot-dashed line).
The CGL prediction is in good agreement for
values of $T_\parallel/T_\perp$ less than about 2, but then
it greatly overestimates the anisotropy (also for the test-particle run). The effect of the firehose instability in
reducing the anisotropy at different distances is clear, and it is not surprising that the anisotropy is higher when the particles are not scattered by electromagnetic fluctuations.
We show in the bottom panel of Figure \ref{anisotropy_super} the ratio between
the values of solid lines and dashed lines. This represents the
reduction of anisotropy due to the presence of fluctuations. One can see that this reduction stands
between 20$\%$ and 40$\%$ depending on location and
time.

\begin{figure}
\center
\includegraphics[width=80mm]{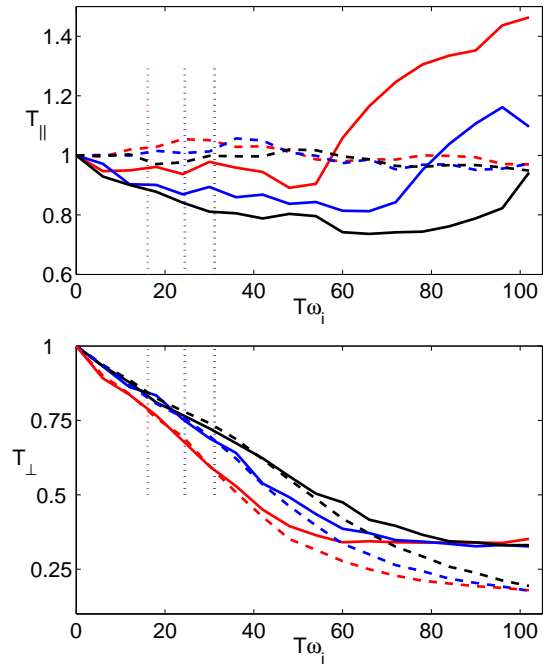}
\caption{Supersonic case: $T_\parallel$ (top panel) and $T_\perp$ (bottom panel). Solid lines are for the self-consistent run, and dashed lines are for  the test-particle run, at radial distances: $r\omega_i/c=1.50$ (red), $r\omega_i/c=1.86$ (blue), $r\omega_i/c=2.22$ (black). The vertical dotted lines indicate the approximate time at which the condition of equation \ref{thre} holds, that is the plasma is marginally stable.}\label{t_par_perp_super}
\end{figure}

We want now to quantify the effect of the instability
by looking separately at the parallel and perpendicular
temperatures. These are shown respectively in the top and
bottom panel of Figure \ref{t_par_perp_super}, where the temperatures have been normalized to their value at time $T=0$. The colours are the same as for
Figure \ref{anisotropy_super}, and solid and dashed lines are again for self-consistent and test-particle runs, respectively. Here we have also pointed, with dotted vertical
lines, the approximate time at which, for different locations,
the local anisotropy and plasma beta are such that
Eq. (\ref{thre}) predicts a marginal stability state. In other words,
after the time denoted for each curve by the dotted line, the plasma is unstable at that location. It has to be reminded
however, that Eq. (\ref{thre}) has been derived from the linear
theory of plasmas in an homogeneous field, and in a periodic
Cartesian geometry. It appears indeed that the curves are
not sensitive to the particular times designed by the
dotted lines. We will give two possible explanations later.
First we comment on the general features that emerge from
Figure \ref{t_par_perp_super}. Recall that the expansion (without the
effect of instabilities) should
result in a monotonic decrease of $T_\perp$, and a constant
$T_\parallel$. This is indeed what happens for the dashed lines.
At a certain time, the value of $T_\perp$ approach an asymptote, and this
happens for successive times at larger distance. This is because
the perpendicular temperature is inversely proportional to the
distance travelled by the particles. This distance will increase
at the beginning of the simulation, until it will reach the maximum
possible value at each location, equal to the distance from the inner boundary, and than will stay constant.
On the other hand, the parallel temperature stays roughly constant, in the test-particle run, while the effect of the instability is supposed to decrease it.
Two comments are in order. First, the stationary
value reached by $T_\perp$  is higher in the self-consistent run, meaning that electromagnetic fluctuations act to reduce the
decrease of perpendicular temperature. However, no particular change is apparent in either $T_\parallel$ or $T_\perp$ when the plasma becomes unstable.
Second, the parallel temperature suffers a mild decrease in the initial stage and then increases at successive times, at larger distance.
This is probably due to the concurrent saturation of firehose modes that once damping tend to enhance the parallel temperature, and
the creation of non-thermal features in the particle distribution function, as we will show later. Since
those modes are convected outwards, the increase in parallel
fluctuation temperature moves outwards too.

\begin{figure}
\center
\includegraphics[width=80mm]{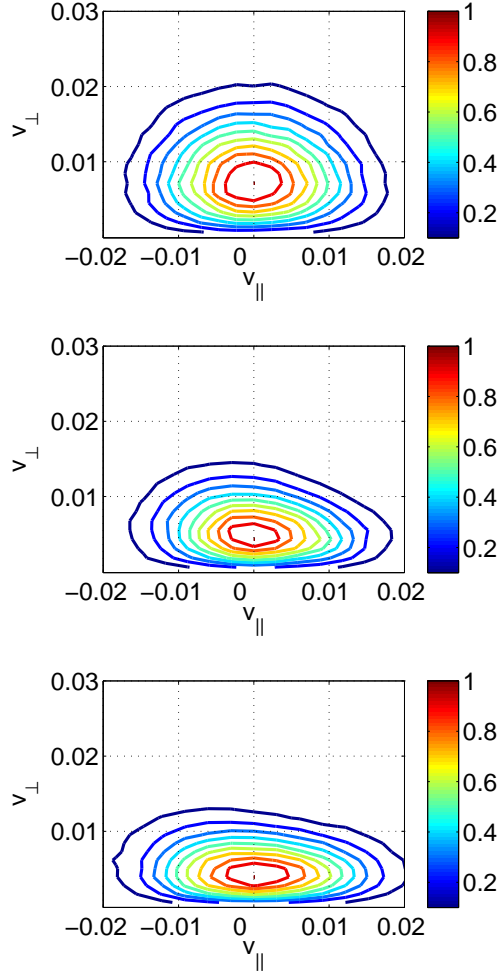}
\caption{Supersonic case. Electron distribution function at times $T\omega_i=18$ (top), $T\omega_i=60$ (middle), $T\omega_i=100$ (bottom), in $(v_\parallel,v_\perp)$ space. The distributions are averaged over the whole active box.}\label{pdf_super}
\end{figure}

What emerges therefore is that the electromagnetic fluctuations are
responsible for slowing down the rate at which the
parallel anisotropy grows. However, we have been unable
to unequivocally identify the firehose instability as
entirely responsible for this behavior.
As we anticipated there are, in our view, two possible
explanations.
One is that the linear theory result summarised in Eq. (\ref{thre})
is not completely applicable when the geometry is
radial, and the magnetic field and density are not
uniform. This is a point that deserves a deeper investigation,
but it could be that the results developed over the years for homogeneous plasma are not straightforwardly applicable to more realistic situations (i.e. for non-constant magnetic field and density, and not periodic structures in the radial direction).
A second possibility is that small noisy fluctuations, unrelated to the firehose instability, could scatter the particles and decrease the parallel anisotropy even before reaching a linearly unstable condition. Moreover, the increase in parallel temperature at later times, seems to be an artificial effect due to the development of non-Maxwellian features in the particle distribution function, such as high energy tails. We show in Figure \ref{pdf_super} the contour plot of the electron distribution function, averaged in the whole active region, at times $T\omega_i=18$ (top panel), $T\omega_i=60$ (middle), $T\omega_i=100$ (bottom). One can notice that as the perpendicular temperature is reduced, the distribution becomes asymmetric in the parallel direction. This is reflected in the increased parallel temperature. The distribution function for states close to the equilibrium is therefore non-bi-Maxwellian, and this is consistent with many satellite observations.

\subsection{Static case: $V_m=0$}
In this section we show the results of one simulation, where the particles have no initial mean velocity ($V_m=0$), but the initial anisotropy $T_\perp/T_\parallel=0.7$. In this way the firehose instability is triggered in the whole active region from the beginning of the simulation. Similar simulations, but for an homogeneous plasma in a double periodic box have been performed in \citet{camporeale08}. The purpose of this run is to check that although the geometry is 2D in space, and therefore some simplifying assumptions have been made on the initial profile of magnetic field and density, and on somehow artificial boundary conditions, the simulations are still able to capture the firehose instability and the consequent decrease of anisotropy. Moreover, with this run we will be able once again to clearly identify the decisive role of the expansion.

\begin{figure}
\center
\includegraphics[width=90mm]{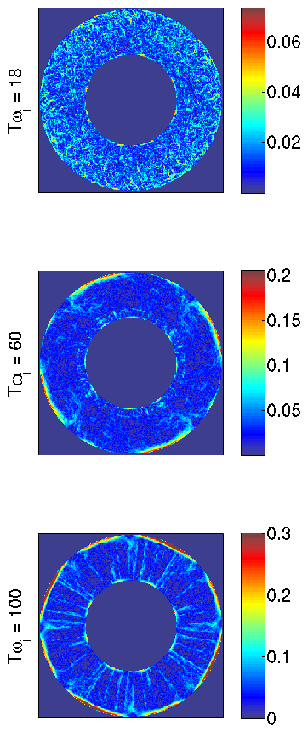}
\caption{Static case: $\delta \mathbf{B}/\mathbf{B_0}$ at times $T\omega_i=18$ (top), $T\omega_i=60$ (middle), $T\omega_i=100$ (bottom). The images are obtained by successive rotations of the box.}\label{delta_B_static}
\end{figure}

\begin{figure}
\center
\includegraphics[width=80mm]{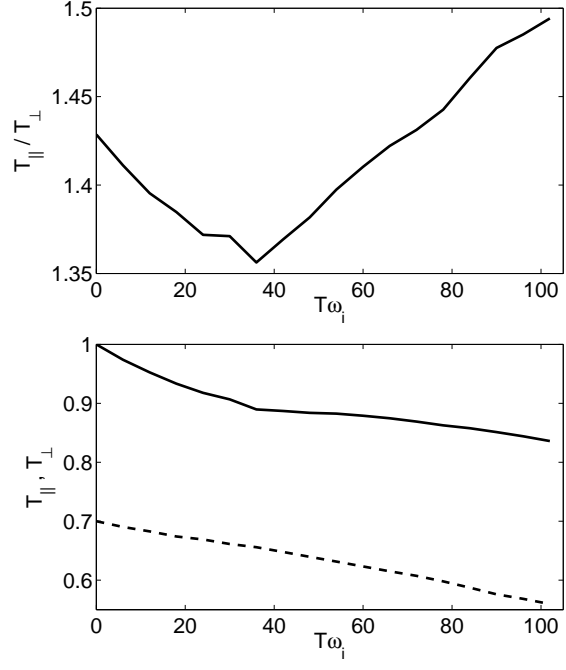}
\caption{Static case. Top panel: anisotropy $T_\parallel/T_\perp$. Bottom panel: $T_\parallel$ (solid line), and $T_\perp$ (dashed line).}\label{t_par_perp_static}
\end{figure}

We show in Figure \ref{delta_B_static} the development of $\delta \mathbf{B}/\mathbf{B_0}$. Here again the dominance of quasi-perpendicular modes at later times is evident. The growth rate of the instability is now higher at larger distances, where the plasma beta is higher. Since we start from an unstable plasma, and the counter-effect of the expansion is now absent, the fluctuations reach amplitudes slightly higher than for the supersonic case (Figure \ref{delta_B_super}). If we look at the temperature anisotropy (Figure \ref{t_par_perp_static}, top panel), and at the parallel and perpendicular temperatures (Figure \ref{t_par_perp_static}, bottom panel, respectively in solid and dashed lines), we see that the anisotropy decreases straight away from the beginning, but not dramatically in value. The temperatures in Figure \ref{t_par_perp_static} are averaged over the whole active box. The decrease of the anisotropy is caused by the fact that the parallel temperature decreases faster than the perpendicular one. At time $T\omega_i\sim 40$, however the parallel temperature reaches a plateau, causing an increase of the anisotropy, because the perpendicular temperature keeps decreasing.

It interesting that the decrease of perpendicular temperature seems to be related to the geometry and the initial profile of magnetic field and density. Even if the initial mean velocity is zero, the magnetic field profile causes a narrowing of the pitch angle distribution for particle moving outwards. Thus the initial conditions favor a distribution of particles that tend to align their velocity with the magnetic field, at the expense of the perpendicular velocity. From Figure \ref{t_par_perp_static} the parallel temperature will decrease if the firehose instability is triggered, but it might be that a more rapid decrease of the perpendicular temperature, when the linear stage of the instability has saturated, will result in a growth of anisotropy. Another interesting point is that the firehose instability, in this configuration, is not as effective in isotropize the particles as is it would be for a non-cylindrical geometry, with constant magnetic field. Indeed, it has been shown in \citet{camporeale08}, that in a Cartesian geometry, an anisotropic plasma is forced by the firehose instability to reduce its anisotropy to a state where the plasma remains close to marginal stability.

\subsection{Subsonic case: $V_m=0.625$}
\begin{figure}
\center
\includegraphics[width=90mm]{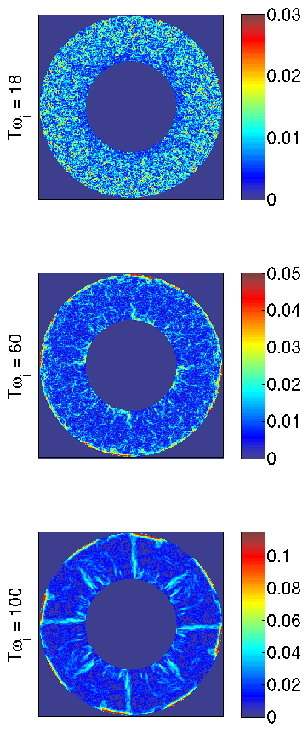}
\caption{Subsonic case: $\delta \mathbf{B}/\mathbf{B_0}$ at times $T\omega_i=18$ (top), $T\omega_i=60$ (middle), $T\omega_i=100$ (bottom). The images are obtained by successive rotations of the box.}\label{delta_B_sub}
\end{figure}

In the subsonic case the interpretation of the results becomes less straightforward, because a vast proportion of electrons are now counter-propagating (i.e moving towards the origin). This results in a non-locality of processes, where particles scattered at one location can rapidly influence the development of instabilities at other locations. Also, the convection of electromagnetic fluctuations towards outer regions is not as efficient as for the supersonic case. Figure \ref{delta_B_sub} shows the development of $\delta \mathbf{B}/\mathbf{B_0}$, as for the previous cases. The whole dynamics is clearly slower, but the results are consistent with Figures \ref{delta_B_super} and \ref{delta_B_static}, with again the formation of structures aligned with the background magnetic field.

\begin{figure}
\center
\epsscale{.80}
\includegraphics[width=80mm]{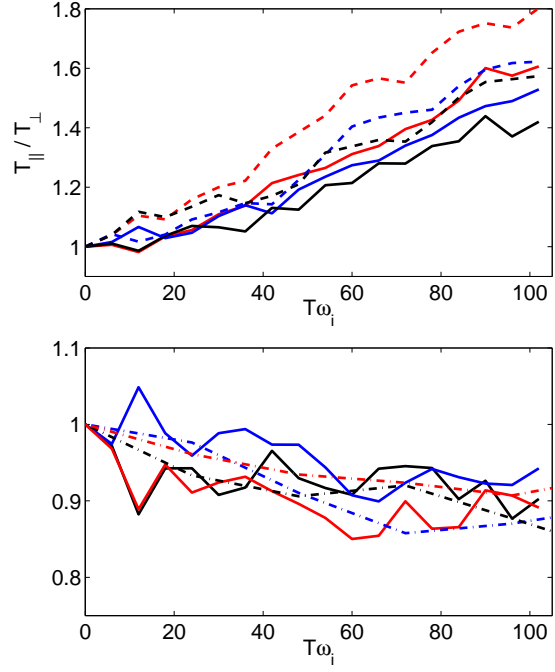}
\caption{Subsonic case. Top panel: anisotropy $T_\parallel/T_\perp$ at radial distances: $r\omega_i/c=1.50$ (red), $r\omega_i/c=1.86$ (blue), $r\omega_i/c=2.22$ (black). Solid lines are for the self-consistent run, and dashed lines are for  the test-particle run. Bottom panel: ratio of anisotropy for self-consistent over test-particle runs. Solid lines are for the subsonic case, and dot-dashed lines are for the supersonic case, with time rescaled by a factor of 4.}\label{anisotropy_sub}
\end{figure}

The top panel of Figure \ref{anisotropy_sub} shows the development in time of the temperature anisotropy (the format and legend are the same as in Figure \ref{anisotropy_super}). These simulations are run again until $T\omega_i=100$, but the results should be compared with the results for the supersonic case, obtained only until  $T\omega_i\sim 25$, since the flow speed is here four times slower. In the lower panel of Figure \ref{anisotropy_sub} we show with solid line the ratio of anisotropy for self-consistent run over test-particle run. The dashed lines show the same quantity, for the supersonic case (i.e. bottom panel of Figure \ref{anisotropy_super}), where now the time has been rescaled by a factor of 4. The two runs, for subsonic and supersonic flows, are qualitatively very similar, with the anisotropy reduced of about 5-15$\%$. This is a sign that the growth rate of the firehose instability for an expanding plasma, must be dependent also on the flow speed of the plasma. Indeed, if this was not the case, the competition of the expansion and the instability would have led to qualitatively different results for the supersonic and subsonic cases.

\section{Conclusions and discussion}
We have presented the results of PIC simulations of a plasma expanding radially on a disc, where the magnetic field and density profile decrease linearly with the inverse of the distance. Although those simulations bear some simplification and assumptions with respect to a realistic stellar wind, they have the unique feature of treating self-consistently the effects due to the expansion and the electromagnetic fluctuations. In fact, we have used a fully-kinetic PIC code, with physical ion-to-electron mass ratio, and a computational box in a fixed frame of reference. Hence, we think that the results of this paper, summarised in this section, might be relevant for the understanding of a more realistic scenario.

We have confirmed that the effect of electromagnetic fluctuations is to decrease the temperature anisotropy: while the expansion makes the parallel anisotropy grow, this increase is slowed down by the presence of EM fluctuations. It is interesting that the presence of fluctuations acts not only in decreasing the parallel temperature, as it is expected, but it also reduces the rate at which the perpendicular temperature decreases during the expansion. The length and time scale constraints of our simulations limit our results to the evolution of the electron temperature anisotropy and its associated instabilities.

The simulations have confirmed the presence of quasi-perpendicular waves, consistent with the development of firehose instability. However, the feedback effect played by fluctuations that counteracts the expansion, does not become more evident, when the plasma becomes linearly unstable, but is rather a continuous effect active since the beginning of the expansion. On the other hand, we have verified that if the expansion would not be present, and the plasma would be injected starting with a parallel anisotropy, this anisotropy would be reduced straight away. The fact that the results are not consistent with linear theory predictions should be thought as a consequence of both the geometry and the fact that the plasma is drifting. Both these effects are neglected in standard linear theory, and although the approximation of a curved geometry with a planar one might be justified at large distances from the Sun, the drift should be probably taken in account. A rough estimate of the importance of the drift, can be made by using the characteristic solar wind parameters listed in \citet{bruno05b}. The leading modes of the electron firehose instability have a wavevector that, depending on the anisotropy, ranges between $kc/\omega_i\sim 20$ to $kc/\omega_i\sim 60$. This corresponds, at 1 AU, to wavelengths of the order of about 10 km. The growth rate is a function of the parallel beta, and the anisotropy, but if we take as a representative value of a fast growing mode a rate of $0.1\Omega_e$ ($\Omega_e$ is the electron cyclotron frequency), then it would take approximately $5\cdot 10^{-2}$ seconds for the wave to grow by a factor of $\mathit{e}$.
The bulk velocity of the solar wind varies between 350 km/s (slow wind) and 600 km/s or more (fast wind), and the electron thermal velocity is between 2000 and 3000 km/s. This means that the electrons can travel a distance comparable, if not greater, than the wavelength of the modes of interest, in a fraction of the growth time.

Moreover, we have shown that the results for subsonic and supersonic flows are qualitatively similar, and the dynamics of the subsonic flow is just slower. This suggests that the growth rate of the electron firehose instability must be a function of the drift speed of the plasma. This is because the expansion and the electromagnetic fluctuations play opposite role for the development of the temperature anisotropy. If the instability growth rate would not be a function of the drift speed, the temperature anisotropy would have been reduced more rapidly in the subsonic case, since here the increase of the anisotropy due to the expansion is slower.

It has to be mentioned that another possible mechanism that is thought to isotropize the distribution function is played by collisions. It has been reported that exists an observational correlation between collisional age and electron temperature anisotropy in the solar wind \citep{salem03}. Clearly, the role of collisions is not included in our simulations.

\acknowledgments
This work was supported by STFC grant PP/E001424/1.

\clearpage



\clearpage







\clearpage

\end{document}